\documentclass[12pt, article]{JHEP3}
\usepackage{amssymb}
\usepackage{verbatim}

\renewcommand{\a}{\alpha}
\renewcommand{\b}{\beta}
\newcommand{\g}{\gamma}     
\newcommand{\del}{\delta}

\newcommand{\m}{\mu}
\newcommand{\n}{\nu}

\renewcommand{\d}{\partial}
\newcommand{\dbar}{\bar{\partial}}

\renewcommand{\det}{\mathrm{det}}
\newcommand{\be}{\begin{equation}}
\newcommand{\ee}{\end{equation}}
\newcommand{\bea}{\begin{eqnarray}}
\newcommand{\eea}{\end{eqnarray}}
\newcommand{\bary}{\begin{array}}
\newcommand{\eary}{\end{array}}

\newcommand{\nn} {\nonumber \\}

\newcommand {\eqr} [1]  {{(\ref{#1})}}

\newcommand{\file}[1]{}
\newcommand{\cS}    {{\mathcal S}}

\newcommand{\tX} {{\tilde X}}

\def\tr {{\mathrm {tr}}}       

\newcommand{\half}  {\frac 1 2}

\newcommand{\bra}{\langle}
\newcommand{\ket}{\rangle}

\preprint{hep-th/0406245}
\keywords{string theory, open string field theory, DBI, D-branes}
\title{Open String Gravity?}
\author{
Yeuk-Kwan E. Cheung, Mark Laidlaw and Konstantin Savvidy\\
Perimeter Institute for Theoretical Physics\\
35 King St. North,  Waterloo, ON N2J  2W9, Canada
}

\abstract{ We present  a new  application of   Boundary String Field
Theory:   calculating the induced-gravity action on a D-brane. Using  a
simple quadratic  tachyon potential to model a D-brane fluctuating  in the
flat target space we derive the  effective action in terms of the
extrinsic curvature to all orders in $ \alpha'$.  We  identify both the
Born-Infeld structure  as well as the Einstein-Hilbert term at  order
$\alpha'$. This corroborates   the conjectured existence of the latter
term in the brane-world scenarios. }

\begin{document}

\section{Introduction}

The brane-world scenario~\cite{Randall:1999ee} -- confining degrees
of freedom within a subspace inside a higher dimensional space -- has
been one of the interesting new ideas in theoretical physics in the past
years.  This scenario has been advocated in attempts to find solutions
to the hierarchy problem, to explain the cosmological constant and 
dark matter, and to parametrize the matter content of our
world.  One theme that receives attention concerns the description of
gravity on the brane, its localization properties and the effects of
higher dimensional gravity.  Many authors have introduced in a
physically motivated manner the Hilbert-Einstein
term and higher curvature invariants to the action in the brane
world-volume.  There is no universal agreement in the literature,
however, about what corrections are appropriate to include.  Due to
the volume of work on this subject, the readers are referred
to~\cite{Arkani-Hamed:1998rs, Antoniadis:1998ig, Gogberashvili:1998vx,
Randall:1999ee, Dvali:2000hr, Kiritsis:2001bc} for a sample of the original works and
to~\cite{Gabadadze:2003ii, Csaki:2004ay} for thorough reviews.  In light
of these exciting developments we  derive
this type of action directly from a fundamentally stringy approach. 
Several authors have taken a related but different approach, and  work out
the world-volume action by examining string amplitudes  for scattering of
gravitons into various world-volume fields~\cite{Green:2000ke,
Corley:2001hg, Ardalan:2002qt, Fotopoulos:2002wy}.

The dynamical degrees of freedom on the brane are those of the
transverse excitations.  These massless scalars, $\phi$, originate
physically from the spontaneously broken translation symmetry of the
position of the brane. In string theory these degrees of freedom are
part of the massless spectrum of the open strings ending on the brane. 
Together with the other massless open strings which are the gauge fields
they can be described by the Dirac-Born-Infeld (DBI)
action~\cite{Fradkin:1985qd, Bergshoeff:1987at, Abouelsaood:1987gd,
Leigh:1989jq}. The standard form of the DBI action incorporates the
higher powers of the gauge field strength (curvature) in fact it does so
to all orders in $\a'$. We observe that the corrections to the D-brane
effective action arising from higher orders in the curvature of the
brane world-volume are relatively much less explored. In this paper we
develop a method to systematically compute in the framework of open
string field theory those corrections to any given order in the
extrinsic curvature of the brane world-volume. This expansion in power
of $\a'$ like in the case of DBI, incorporates stringy effects of only
worldsheets of the lowest genus, the disk.


To an observer living on the worldvolume of the flat brane (embedded in
flat ambient space) the transverse deviations principally manifest
themselves
as perturbations of the metric away from flat Minkowski metric.
$$
G_{\m\n} = \eta_{\m\n} + \d_\m \phi \, \d_\nu \phi
$$
 These metric perturbations
propagate at the speed of light and may conceivably be detected by an apparatus
like LIGO.  In this respect those metric perturbations are acting similarly to
gravity.  However we would like to stress some important differences:
\begin{itemize}
\item{
The number of dynamical degrees of freedom is equal to the co-dimension
of the brane and is independent of the dimensionality of the worldvolume.
}
\item{
The degrees of freedom of this induced gravity get their kinetic term not
from the Einstein-Hilbert term, present also in this case, but from the brane tension. }
\end{itemize}
These are in stark contrast to ordinary gravity, or indeed the
brane-gravity approach where the entries of the metric tensor are the 
degrees of freedom and Einstein-Hilbert term provides the kinetic term for the graviton
and the tension term would be interpreted as the cosmological constant.
Ultimately, the question of whether or not the present theory reproduces
the known facts about gravitational force is likely to hinge  on the
interaction of the metric perturbations with matter. On this front,
there is the reassuring evidence that the photon field, representing
the open-string gauge field degrees of freedom, does in fact have  the 
required coupling through the energy-momentum tensor to the said
metric perturbations.

In this letter we propose a simple technology, based on boundary
string field theory, for the computation of the higher curvature
corrections to the induced action on a D-brane.
For the sake of simplicity we are setting $F=0$ and so  concern ourselves 
 with only the induced metric degrees of freedom.
To model a D-brane we insert  a  quadratic tachyon potential on the boundary
of the string worldsheet to confine the endpoints of the open string.
We obtain a closed expression to all order of $\a'$:
$$
\cS = \int dx_{||} \, \sqrt{\det( G_{\m\n} )} ~ 
      \int dx_{\perp} ~ e^{-x_{\perp}^2}  \exp \left( \sum_j (-)^j \frac {\zeta(j)}{j} \tr (l_s \, K \cdot x_{\perp})^j \right)
$$
where $K$ is the extrinsic curvature.

  Our approach has a number of
advantages, the principal one being its calculational simplicity.
 A novel point is that
we do not introduce the position of the brane under consideration as part
of a gauge field, but rather encode it in the open string tachyon.
This allows us to progress with considerably less effort past the Born-Infeld limit.
As we shall explain below, we base our analysis only on the properties of the quadratic
interactions on the boundaries of an open string worldsheet.
 We exhibit our method for the bosonic string but it can easily be  adapted to its
 supersymmetric cousins because the  BSFT
for the superstrings have been developed~\cite{Niarchos:2001si,Kutasov:2000aq}.
One can also generalize to include higher drivative terms not present
in this work.

\section{The Model}

Our approach is based on the old ideas of  string coupling in background
 fields~\cite{Fradkin:1985qd, Abouelsaood:1987gd}
and the boundary string field theory
formalism~\cite{Witten:1992qy, Witten:1993cr, Shatashvili:1993kk}:
We couple some terms to the boundary of the string
world-sheet, and calculate the  partition function of the two dimensional
sigma model for these fields.  We then identify this partition functional
of world-sheet couplings as a space-time action for these couplings,
interpreted as fields.  This approach was used originally to calculate the
Born-Infeld action and string  loop corrections for a gauge field, and was
formalized later  for massive and tachyonic boundary interactions.

We consider a toy model of a Dp-brane with world-sheet action on the disk
\be
S = \frac1{4\pi \a'} \int_{\Sigma} \d X^M \, \dbar X_M + 
	\frac{c}{4\pi \a' } \int_{\d \Sigma} 
	\left( X^i - Y^i(X^\mu) \right)^2 ~,
\label{eq:stringS}
\ee
where $\Sigma$ and $\d\Sigma$ are respectively the worldsheet and its 
boundary.
The brane is parametrized in the static gauge by $p+1$ coordinates $X^\m$, while
being positioned at $Y^i$ in the remaining $d-p+1$ directions.
Taking the limit $c \rightarrow \infty$  will ensure that the endpoint of the string
are confined to move on the hypersurface defined by $Y^i(X^\mu)$.
We use capital Roman indices $M,N$ to denote the total target 
space, if needed.


We compute the partition functional of $c$ and $Y^i$ 
through a Euclidean  path integral
\be \label{eq:stringZ}
Z(c,Y^i) = \int dX^i \, dX^\mu \, e^{-S} ~.
\ee
As was discussed in~\cite{Witten:1993cr} for constant $Y^i$ there is a 
smooth interpolation between 
Neumann and Dirichlet boundary conditions for the coordinates $X^i$ as $c$ 
goes between $0$ and $\infty$.
In particular, the limit $c\rightarrow \infty$ is a conformally invariant 
point of the theory and we identify the partition 
functional in this limit with a spacetime effective action for the $Y^i$s.  
The general relation in bosonic BSFT is
\be
S = \left( 1 + \beta_i \, \d_i \right) Z~,
\ee
 where the derivatives are with respect to the boundary couplings and the
$\beta$s are the corresponding beta-functions.  
In this case we consider the $c=\infty$ limit, which is a zero of the 
tachyon beta-function, and the fields $Y$ break the translational 
symmetry in the directions $X^i$ and so are massless.  We then can  identify
\be 
\lim_{c \rightarrow \infty} Z(c,Y^i) = S(Y^i) ~.
\ee

Prior to calculating the partition functional associated with
the action \eqr{eq:stringS} we review some useful facts concerning quadratic
boundary interactions.  
For a single boson $X$ with boundary interaction $c \oint  X^2$,
the propagator for the oscillatory modes on the unit disk is 
\bea
\langle X(z) \, X(z') \rangle = \half \, \sum_{n=1}^\infty 
 \frac{ (z \, \bar z')^n + (\bar z \, z')^n}{n+ 2\, c}~.
\label{eq:XX}
\eea
There is also an associated zero mode on the world sheet.
One can show~{\cite{Witten:1993cr,Kraus:2000nj}}
that the partition function
for this boson (including the zero mode) is given by
\bea \label{eq:singleu}
Z &=& \int dx \, e^{-c x^2} \, \prod_{n=1}^\infty \frac{1}{1 + 2 \, c/n} \nn
&=&   \int dx \, e^{-c x^2} \, \exp \left( \sum_k (-)^k \frac {\zeta(k)}{k} (2 \, c)^k \right) \nn
&=&   \int dx \, e^{-c x^2} \, e^{2 a  c} ~ \Gamma(1 + 2 \, c)~.
\eea
where $x$ denotes the zero mode of the string field.
The parameter $a$ is an undetermined coefficient that regularizes
the correlator $ \bra X^2(0)\ket $.
Technically it is associated with the $\zeta$-function representation
of the infinite product:
$a = \zeta(1)+\gamma$ can assume different values depending on the regularization scheme
used, however it does not affect physical quantities such as
the ratio of brane tensions {\cite{Kraus:2000nj}}.
It is straightforward to generalize these results to multiple  $X$'s
with arbitrary quadratic interactions {\cite{Laidlaw:2001jt}}.

We now wish to write the boundary interaction term of \eqr{eq:stringS} as
a sum of a quadratic term and terms cubic and higher.
We first write the string fields $X$ in terms of zero mode and
oscillators:
$$
	X= x + \tX
$$
and  Taylor expand $Y(X^\mu)$ around the
classical position of the brane, $Y^i(x)$, in powers of $\tX$:
\bea
Y^i(X^\mu) = Y^i(x^\mu) + \d_\nu Y^i(x^\mu) \, \tX^\nu + \frac{1}{2} \d_{\nu}
\d_{\alpha} Y^i(x^\mu) \tX^\nu \, \tX^\alpha + \ldots
\eea
Having done this the interaction term of~\eqr{eq:stringS} becomes:
\bea \label{eq:Iexpanded}
S_{int} &\equiv & c \int_{\d \Sigma}(x^i -Y^i)^2 +  \tX^M U_{MN} \tX^N + ...\nn
&=&c \int_{\d \Sigma} (x^i -Y^i)^2 + (\tX^i-\d_\mu  Y^i\tX^\mu - \half\d_{\mu\nu} Y^i \tX^\mu \tX^\nu)^2  \nn
&& ~ +2 (x^i -Y^i) (\tX^i -\d_\mu Y^i\tX^\mu -\half\d_{\mu\nu}Y^i \tX^\mu \tX^\nu) +\cdots~~
\eea
where we can read off $U$
\be \label{eq:U}
 U_{MN} = \left( \begin{array}{cc}
                   \delta^{ij} &  -\d_\nu Y^i \\
		   -\d_\mu Y^j &  \d_\mu Y^i \d_\nu Y^i -(x^i -Y^i)\d_{\mu\nu} Y^i
		   \end{array}  
	  \right)
\ee 
We note that the  boundary term linear in $\tX$ does not contribute
to the partition function since the propagator \eqr{eq:XX} is periodic on the
boundary of the unit disk.
Furthermore since the propagator is proportional to $1/c$ in the limit
$c\rightarrow\infty$,
we can neglect cubic and higher boundary terms in
the interaction  because they give rise to $\frac{1}{c}$ corrections.

The form of the interaction   calls for a field redefinition, $\tX \rightarrow J^{-1} \tX$:
\bea  \label{eq:fieldredef}
 \tX^i &\rightarrow& \tX^i + \d_\mu Y^i \tX^\mu  \nn
 \tX^\mu  &\rightarrow& \tX^\mu - \d_\mu Y^i \tX^i
\eea
and 
\be \label{eq:newU}
  U \rightarrow J^T\, U\, J
\ee
where $J^T$ is the transpose of $J$.
%
Note also that the new metric
\be \label{eq:newG}
G = J J^T
\ee 
is again block diagonal, while its longitudinal part coincides with the induced metric.  
Therefore the transformed fields $X^i$ are 
locally perpendicular  and $X^\mu$ are locally parallel to the brane. 

The above field redefinition gives rise to a Jacobian for each oscillator mode.
The Jacobian is
\bea
\det \left(J^{-1}\right) = \left| \begin{array}{cc} \delta^{ij} & - \d^\mu
Y^i \\
\d^\nu Y^j & \delta^{\mu\nu} \end{array} \right|,
\eea
so by $\zeta-$function regularization,
\bea
\prod_{n=1}^\infty \det J &=& \left(\det J\right)^{\zeta(0)} = \left(\det
J\right)^{-1/2}.
\eea

We now follow \cite{Witten:1993cr,Kraus:2000nj} and integrate out $\tX$
from
\eqr{eq:stringZ} using the techniques outlined in \eqr{eq:XX} and \eqr{eq:singleu}.  
The partition functional is then naturally divided into contributions from
the directions transverse to the world-volume, and the directions along
the world-volume.
The partition functional is
\bea  \label{eq:zmZ}
Z(c,Y^i)
&=& \int dx^M e^{-c (x^i - Y^i)^2} (\det J)^{-\half}
\exp \left( \sum_k (-)^k \frac {\zeta(k)}{k} (2 \, c \, (U G)_{MN})^k \right) \nn
&=& Z(c) \int dx^M e^{-c (x^i - Y^i)^2} (\det J)^{-\half} e^{2ac (UG)^\mu_\mu }
\det \Gamma\left( 1 + 2  c (U G)_{\mu\nu} \right) \nn
\eea
We use $ \tr (J^T U J)^k = \tr (UG)^k$, with $G$ defined by~\eqr{eq:newG}. 
In the above expression we
have defined $Z(c)$ to be the partition function of the 
transverse modes.

We now wish to integrate out the (world-sheet) zero modes transverse to
the brane.  
Our strategy will be to Taylor expand~\eqr{eq:zmZ} in $x-Y$ and perform 
the Gaussian integral order by order in this expansion.  Since the 
function is strongly peaked around $x^i = Y^i$ the portion of 
$U G$  which depends on $x-Y$ is an appropriate  small parameter.  
Motivated by desire to write the result in covariant form\footnote{
We recall a brief calculation in differential geometry.
Consider an induced metric of the form
\be
g_{\mu\nu} = \eta_{\mu\nu} + \d_\mu \phi^i(x^\a) \d_\nu \phi^i(x^\a)~.\nonumber
\ee
The Ricci tensor for this metric is
\be
R_{\nu\b} = g^{\mu\a} \left(\d_\mu \d_\a \phi^i \d_\nu \d_\b \phi^j
- \d_\a \d_\b \phi^i \d_\mu \d_\n \phi^j \right) \left( \delta^{ij} -
g^{\del \g} \d_\del \phi^j \d_\g \phi^i \right) ~.\nonumber
\ee
Similarly, with the index $i$ perpendicular to the $\mu,\nu$, the 
extrinsic curvature is
\be
K_{\mu\nu}^i = \d_\mu \d_\nu \phi^i ~. \nonumber
\ee
}
we introduce the extrinsic curvature $K^i_{\mu\nu} = \d_{\mu}\d_{\nu}Y^i$.
When we Taylor expand~\eqr{eq:zmZ} in $(x^i - Y^i) K^i_{\mu\nu}$ we obtain
\bea
Z(c,Y) &=& Z(c) \int dx^M e^{-c (x^i - Y^i)^2} (\det J)^{-\half} \nn
&& ~\Bigg\{  1 + 2 \bigg( a^2 \tr K^i \tr K^j + \zeta(2) \tr K^i K^j 
 	\bigg) (x^i-Y^i)(x^j-Y^j) \nn
&& ~~~ + \frac23 \bigg( a^4 \tr K^i \tr K^j \tr K^l \tr K^m 
       + 6 a^2 \zeta(2) \tr K^i \tr K^j \tr K^l K^m \nn 
&& ~~~ +  6 \zeta(4) \tr K^i K^j K^l K^m
       - 8 a \zeta(3) \tr K^i \tr K^j K^l K^m  \nn
&& ~~~       + 3\zeta(2)^2 \tr(K^i K^j) \tr(K^i K^j) \bigg)
         (x^i-Y^i)(x^j-Y^j) (x^l-Y^l)(x^m-Y^m) + \cdots \Bigg\} \nn
	 &&
\eea
  
We note that the choice of $a^2 = -\zeta(2)$ results
in the $O(\a')$ terms being expressed in a simple way, however, 
this corresponds to setting the $z \rightarrow z'$ limit of 
\eqr{eq:XX} to be $i \sqrt{\zeta(2)}$, a somewhat non-standard 
regularization. 
We restore the dependence on $\a'$, and simultaneously note that
there is no longer any dependence on the parameter $c$, at every order in the expansion below.
We perform
the integrations for the terms both quadratic and quartic in $x^i-Y^i$ and
obtain an effective action for the fluctuations $Y^i$,
\bea
S(Y) &=& \mathrm{Z(c)} \int dx^\mu \sqrt{\det(G_{\m\n})}
   \Bigg( 1 - \a' \zeta(2) R +  {\a'}^2 \zeta(2)^2 \frac{1}{2} R^2 \nonumber \\ 
&& ~~~ +  {\a'}^2 \zeta(2)^2
\left( - 2 \tr K^i K^j \tr K^i \tr K^j + \tr K^i K^j \tr K^i K^j \right) \nonumber \\ 
&& ~~~ + {\a'}^2 \left( -i 4 \sqrt{\zeta(2)} \zeta(3) \right)
\tr K^i \tr K^i K^j K^j  \nonumber \\ 
&& ~~~ +  {\a'}^2 \zeta(4) \left( 2 \tr K^i K^i K^j
K^j +\tr K^i K^j K^i K^j \right) \ldots \Bigg)~.
\label{eq:SY}
\eea
All inner products are with respect to the metric $G_{\mu\nu} =
\eta_{\mu\nu} + \d_\mu Y^i \d_\nu Y^i$.
When we compute~\eqr{eq:SY} we make the identification $(\det J_{MN})^{-1/2} =
\sqrt{\det G_{\m\n}}$ as can be easily verified.
The terms at order ${\a'}^2$ do not either obviously cancel, or combine
into a Gauss-Bonnet term $R^2 + 4 R_{AB}^2 + R_{ABCD}^2$, although it is
certainly possible that some field redefinition could combine these
terms.  We can argue that the imaginary part (or more weakly the presence
of a quadratic term in $R$ indicating that the ghosts do not decouple) is
a reflection of the well-known instability for bosonic branes.

It is also possible, and perhaps more democratic, to write 
\eqr{eq:SY} solely in terms of the extrinsic curvature. We may additionally
discard the indeterminate terms involving $a$, in line with the regularization 
adopted in~\cite{Witten:1992qy}:
\bea
S(Y) &=& \int dx^\mu \sqrt{\det(G_{\m\n})} 
\Bigg[ 1 + 
\a' \, \frac { \pi^2} {6} \, \tr K^i K^i \Bigg. \nn
&+& \a'^2 \, \frac { \pi^4} {18} \, \left( \frac 1 5 \, \left( 2 \, \tr K^i K^i K^j K^j +\tr K^i K^j K^i K^j \right) \right. \nn
&+& \left. \frac 1 4 \, \Bigg. \left(2 \, \tr K^i K^j \, \tr K^i K^j + \tr K^i K^i \, \tr K^j K^j \right) \right) + ... \Bigg]~.
\eea

As an exercise, it is possible to add the coupling to a $U(1)$ gauge field
to this model.  To effect this, we add the coupling
\bea
\del S = \oint F_{\mu\nu} X^\mu \d_t X^\nu
\eea
to the action \eqr{eq:stringS}.
It is very easy to verify that this addition changes the
action \eqr{eq:SY} in the following ways:  First the prefactor which was
the square root of the determinant of the metric becomes the Born-Infeld
term,
\bea
\sqrt{ \det G} \rightarrow \sqrt{\det G + 2 \a' F },
\eea
 and second, the metric used in the contractions of $R$ and $K$ gains a
non-diagonal term
\bea
G^{\mu\nu} &=& G^{\mu \a} \left( \frac{1}{1 + 2 \a' F_{\a \b} G^{\b \nu} }
\right)_\a^\nu.
\eea

An alternative way to derive all the results discussed here is to start
with action~\eqr{eq:stringS}, then Taylor expand the interaction term as
in \eqr{eq:Iexpanded}.  Next impose that the $X^i$ coordinates have
Dirichlet boundary conditions, and treat the $\d_\mu Y^i \d_\nu Y^i$ term
as a boundary mass term.  Calculations exact in this, and perturbative in
$(x^i -Y^i) \d_\mu \d_\nu Y^i$ give the same results as those discussed
above.

We have given this argument in the simplest possible case, namely we take
the limit where $c\rightarrow \infty$.  Using the boundary string field
theory technology it is possible to generalize this construction to the
case of large but finite $c$.  This would describe branes of thickness
$\propto \frac{1}{\sqrt{c}}$ which have been of interest 
(see for example \cite{Csaki:2000fc}).
Also, we have only considered the cases of a bosonic brane with or without
a $U(1)$ gauge field, and the generalization to other types of matter is
of interest.

\section{Discussion}

We now summarize the steps that we took in the derivation of this result.
We started by modelling a hypersurface  by a confining quadratic potential for
the ends of the open string.  The classical position of this surface depended
on the transverse coordinates, and  the limit where the strength of the
potential went to infinity corresponded to Dirichlet boundary conditions
for the open string, and hence the surface was identified as a D-brane.
We then taylor-expanded the boundary interaction term, keeping only the
 quadratic part of the boundary potential, since the cubic and higher
 order terms gave corrections that vanished in the limit we consider.
 A change of variables
so that the $X$ fields were locally tangential or normal to the brane was
performed, and the Jacobian for this transformation was identified with
$\sqrt{ \det(\eta + \d Y \d Y)}$ exactly as we expect for the induced
gravity.
 Then we derived the partition function for the system, and taylor
expanded in the zero mode term, finally integrated out the transverse zero
modes.  This procedure gives us the $\a'$ expansion.  We identified the
result of this integration as the space-time effective action for the
brane, and  expressed  it in terms of curvature invariants on the brane.

The normal ordering prescription that was used to obtain \eqr{eq:SY}
may not be the most natural. We expect that this ambiguity in $a$
will be eliminated in superstrings\footnote{
We would like to thank Ignatios Antoniadis for pointing this out
\cite{Bachas:1999um, Antoniadis:2002tr}.}.
Expressing the action for the transverse
scalars in terms of the extrinsic curvature $K$ may make the resulting
world-volume action more amenable to supersymmetrization%
\footnote{We thank Jerome Gauntlett for making this point to us.}. 
The physics of a brane model with gravity built from the extrinsic
curvature remains an open problem.  Other researchers have considered
theories with a maximum acceleration and found that the generic
Lagrangian in those cases is a power series in the trace of the
Riemann tensor \cite{Schuller:2004rn}.  The present simple model does
not give rise to such a class of theories.  However the nice form into
which the extrinsic and intrinsic curvature have organized themselves
calls for further explanation.
Finally it is very interesting to study the possible implications
of this modified action  to Newtonian gravity.

\section*{Acknowledgement}
We would like to thank Ignatios Antoniadis, Jerome Gauntlett,
Rob Myers, Nemani Suryanarayana,
Frederic Schuller and  Mattias Wohlfarth  for useful  discussions.
ML's research is supported in part by a fellowship from NSERC.
The research at Perimeter Institute is supported by NSERC.
EC and KS would also like to thank the Theory Division of CERN for hospitality where
the final stage of the work was done.

\clearpage
\addcontentsline{toc}{section}{References}
\bibliographystyle{unsrt}
\bibliography{cls}

\begin{thebibliography}{10}

\bibitem{Randall:1999ee}
Lisa Randall and Raman Sundrum.
\newblock A large mass hierarchy from a small extra dimension.
\newblock {\em Phys. Rev. Lett.}, 83:3370--3373, 1999.

\bibitem{Arkani-Hamed:1998rs}
Nima Arkani-Hamed, Savas Dimopoulos, and G.~R. Dvali.
\newblock The hierarchy problem and new dimensions at a millimeter.
\newblock {\em Phys. Lett.}, B429:263--272, 1998.

\bibitem{Antoniadis:1998ig}
Ignatios Antoniadis, Nima Arkani-Hamed, Savas Dimopoulos, and G.~R. Dvali.
\newblock New dimensions at a millimeter to a fermi and superstrings at a tev.
\newblock {\em Phys. Lett.}, B436:257--263, 1998.

\bibitem{Gogberashvili:1998vx}
Merab Gogberashvili.
\newblock Hierarchy problem in the shell-universe model.
\newblock {\em Int. J. Mod. Phys.}, D11:1635--1638, 2002.

\bibitem{Dvali:2000hr}
G.~R. Dvali, Gregory Gabadadze, and Massimo Porrati.
\newblock 4d gravity on a brane in 5d minkowski space.
\newblock {\em Phys. Lett.}, B485:208--214, 2000.

\bibitem{Kiritsis:2001bc}
E.~Kiritsis, N.~Tetradis, and T.~N. Tomaras.
\newblock Thick branes and 4d gravity.
\newblock {\em JHEP}, 08:012, 2001.

\bibitem{Gabadadze:2003ii}
Gregory Gabadadze.
\newblock Ictp lectures on large extra dimensions.
\newblock 2003.

\bibitem{Csaki:2004ay}
Csaba Csaki.
\newblock Tasi lectures on extra dimensions and branes.
\newblock 2004.

\bibitem{Green:2000ke}
Michael~B. Green and Michael Gutperle.
\newblock D-instanton induced interactions on a d3-brane.
\newblock {\em JHEP}, 02:014, 2000.

\bibitem{Corley:2001hg}
Steven Corley, David~A. Lowe, and Sanjaye Ramgoolam.
\newblock Einstein-hilbert action on the brane for the bulk graviton.
\newblock {\em JHEP}, 07:030, 2001.

\bibitem{Ardalan:2002qt}
F.~Ardalan, H.~Arfaei, M.~R. Garousi, and A.~Ghodsi.
\newblock Gravity on noncommutative d-branes.
\newblock {\em Int. J. Mod. Phys.}, A18:1051--1066, 2003.

\bibitem{Fotopoulos:2002wy}
A.~Fotopoulos and A.~A. Tseytlin.
\newblock On gravitational couplings in d-brane action.
\newblock {\em JHEP}, 12:001, 2002.

\bibitem{Fradkin:1985qd}
E.~S. Fradkin and A.~A. Tseytlin.
\newblock Nonlinear electrodynamics from quantized strings.
\newblock {\em Phys. Lett.}, B163:123, 1985.

\bibitem{Bergshoeff:1987at}
E.~Bergshoeff, E.~Sezgin, C.~N. Pope, and P.~K. Townsend.
\newblock The born-infeld action from conformal invariance of the open
  superstring.
\newblock {\em Phys. Lett.}, 188B:70, 1987.

\bibitem{Abouelsaood:1987gd}
A.~Abouelsaood, Jr. Callan, Curtis~G., C.~R. Nappi, and S.~A. Yost.
\newblock Open strings in background gauge fields.
\newblock {\em Nucl. Phys.}, B280:599, 1987.

\bibitem{Leigh:1989jq}
R.~G. Leigh.
\newblock Dirac-born-infeld action from dirichlet sigma model.
\newblock {\em Mod. Phys. Lett.}, A4:2767, 1989.

\bibitem{Niarchos:2001si}
Vasilis Niarchos and Nikolaos Prezas.
\newblock Boundary superstring field theory.
\newblock {\em Nucl. Phys.}, B619:51--74, 2001.
\newblock hep-th/0103102.

\bibitem{Kutasov:2000aq}
David Kutasov, Marcos Marino, and Gregory~W. Moore.
\newblock Remarks on tachyon condensation in superstring field theory.
\newblock 2000.

\bibitem{Witten:1992qy}
Edward Witten.
\newblock On background independent open string field theory.
\newblock {\em Phys. Rev.}, D46:5467--5473, 1992.
\newblock hep-th/9208027.

\bibitem{Witten:1993cr}
Edward Witten.
\newblock Some computations in background independent off-shell string theory.
\newblock {\em Phys. Rev.}, D47:3405--3410, 1993.
\newblock hep-th/9210065.

\bibitem{Shatashvili:1993kk}
Samson~L. Shatashvili.
\newblock Comment on the background independent open string theory.
\newblock {\em Phys. Lett.}, B311:83--86, 1993.
\newblock hep-th/9303143.

\bibitem{Kraus:2000nj}
Per Kraus and Finn Larsen.
\newblock Boundary string field theory of the dd-bar system.
\newblock {\em Phys. Rev.}, D63:106004, 2001.
\newblock hep-th/0012198.

\bibitem{Laidlaw:2001jt}
M.~Laidlaw and G.~W. Semenoff.
\newblock The boundary state formalism and conformal invariance in off-shell
  string theory.
\newblock {\em JHEP}, 11:021, 2003.
\newblock hep-th/0112203.

\bibitem{Csaki:2000fc}
Csaba Csaki, Joshua Erlich, Timothy~J. Hollowood, and Yuri Shirman.
\newblock Universal aspects of gravity localized on thick branes.
\newblock {\em Nucl. Phys.}, B581:309--338, 2000.

\bibitem{Bachas:1999um}
Constantin~P. Bachas, Pascal Bain, and Michael~B. Green.
\newblock Curvature terms in d-brane actions and their m-theory origin.
\newblock {\em JHEP}, 05:011, 1999.

\bibitem{Antoniadis:2002tr}
Ignatios Antoniadis, Ruben Minasian, and Pierre Vanhove.
\newblock Non-compact calabi-yau manifolds and localized gravity.
\newblock {\em Nucl. Phys.}, B648:69--93, 2003.

\bibitem{Schuller:2004rn}
Frederic~P. Schuller and Mattias N.~R. Wohlfarth.
\newblock Sectional curvature bounds in gravity: Regularisation of the
  schwarzschild solution.
\newblock 2004.

\end{thebibliography}

\end{document}